# Coordinated adrenergic and cholinergic neuromodulation facilitate flexible and reliable cortical states that track pupillary fluctuations

Brandon R. Munn[1,2,*], Eli J. Müller[1,2], Gabriel Wainstein[1], and James M. Shine[1].

**Affiliations**

[1]Brain and Mind Centre, The University of Sydney, Sydney, New South Wales, Australia

[2]Complex Systems, School of Physics, The University of Sydney, Sydney, New South Wales, Australia.

[*] Correspondence to: brandon.munn@sydney.edu.au.

**Abstract**

The cholinergic and adrenergic neuromodulatory systems have been linked to focused and exploratory attentional states. Resolving how neuronal activity can be adaptively transitioned to facilitate different behavioural states remains an open question in neuroscience, due to difficulties in tracking neuromodulatory tone and neural recordings simultaneously. Here, we used high-resolution neural recordings and biophysical modelling to develop a sensitive, non-invasive signature for tracking the balance between adrenergic and cholinergic tone from fluctuations in pupil diameter. We utilised this signature in murine spiking and ECoG recordings and found that neuromodulation mediated unique neural dynamics. We then developed a biophysical model of nonlinear layer V pyramidal neurons – which have been shown to be primarily involved in arousal. The model revealed cholinergic (adrenergic) neuromodulation quenches (enhances) neuronal variability supporting reliability (flexibility). Our results establish that different arms of the ascending arousal system facilitate distinct modes of adaptive information processing in the brain.

**Introduction**

Neural activity flexibly supports distinct information processing modes, while retaining the ability to respond rapidly with a high degree of specificity to unique ethological settings. However, it is difficult to conceive how these flexible, yet robust patterns could arise from the relatively fixed structural connections between specialised regions of the brain. By dynamically altering the excitability and receptivity of targeted neurons [1], the ascending neuromodulatory system is ideally placed to mediate precisely the flexibility required to mediate complex, adaptive behaviour.

Adrenergic and cholinergic neuromodulation are two dominant arms of the ascending arousal system, and their activation has been linked to distinct cognitive behavioural states. Specifically, the cholinergic system is often associated with attentional focus and selection [2], whereas the noradrenergic system with exploration and more flexible behaviour [3]. Recent empirical evidence has linked these distinct behavioural states to differential neural dynamics in human fMRI studies, however, this study was limited to analysing BOLD dynamics [4]. As such it remains to be shown how the underlying population neuronal spiking activity can adaptively facilitate differential behavioural states. Anatomically, adrenergic neuromodulation is mediated by diffuse projections from the locus coeruleus (LC; Fig. 1A rose) in the brainstem, whereas cholinergic neuromodulation is mediated cortically by targeted projections from the nucleus basalis of Meynert (nbM) in the basal forebrain, and to the thalamus and midbrain by projections from the pedunculopontine (PPN) and laterodorsal tegmentum within the brainstem (LDT; Fig. 1A blue) [5–8]. While these arousal systems can operate independently, they can also be coupled together following phasic LC bursts that excite the cholinergic hubs, both directly (via $G_q$-mediated augmentation of cholinergic fibres) and indirectly (via $G_i$-mediated inhibition of GABAergic interneurons) [1]. Nevertheless, it is unclear how these macroscale differences in anatomical projections are translated into differential cortical dynamics at the neuronal network scale.

A major site of action for acetylcholine (ACh) and noradrenaline (NAd) in the cerebral cortex are thick-tufted Layer V pyramidal neurons (LV$_{PN}$) [5,6,27]. LV$_{PN}$ are the main output channel from the cerebral cortex and possess two dendritic compartments – somatic and apical – that are electrotonically separated at rest by hyperpolarisation-activated cyclic nucleotide-gated (HCN) channels located along the dendritic apical trunk (Fig. 1B). This electrotonic separation ensures that inputs to the apical dendrites do not typically affect spiking dynamics in the soma, which are instead driven solely by inputs close to the soma [9,10]. However, when the electrotonic separation is overcome, and the dendritic compartments are coupled – e.g., via closure of HCN channels, cortical feedback into layer I, or non-specific thalamic drive – calcium spikes in the apical layer can propagate down the apical trunk and if coincident with somatic sodium spikes generate high-frequency burst firing (Fig. 1C) [5,11]. In this way, LV$_{PN}$ spiking dynamics are nonlinearly dependent upon both the state of the cell and the broader network [12]. Recent studies have confirmed that integral functions of the brain – such as conscious awareness and binding feedforward and feedback signals [5,13] – are coordinated through LV$_{PN}$ bursting. Crucially, ACh and NAd have been shown to modulate LV$_{PN}$ at the apical tufts and the closure of this HCN channel, respectively, increasing the likelihood of burst-firing in LV$_{PN}$ [5–7]. This dependence suggests that neuromodulatory control over LV$_{PN}$ may mediate computationally beneficial adaptive information processing modes, however evidence for this phenomenon is relatively scant.

Here, we use a combination of 2-photon imaging, pupillometry, electrophysiology, and biophysical modelling to test the hypothesis that different arms of the ascending arousal system controllably mediate distinct nonlinear processing modes in the brain. Through this process, we demonstrate that: the balance between NAd and ACh tone can be noninvasively inferred through non-luminance changes in pupil diameter; LC and nbM projection systems mediate flexible and reliable systems-level dynamical modes, respectively; and that the synergistic coupling between NAd/ACh following large phasic bursts of the LC allows the brain to reach novel brain-states. Finally, our analyses establish a mechanistic model of the effects of the ascending arousal system and its capacity to facilitate distinct adaptive information processing modes within the brain.

**Results**
**The balance between adrenergic and cholinergic tone can be non-invasively inferred from pupil fluctuations**
Without access to microdialysis recordings, indirect measures are required to measure changes in neuromodulatory tone. For example, previous work has performed two-photon imaging of the efferent cortical projections of both LC and nbM to infer the effects of neuromodulation on brain state dynamics [14]. Others have relied on an even more indirect measure – namely, the dilation of the pupil [14–16]. Although there are numerous other structures – both cortical and subcortical [17–20] – that modulate pupil diameter, there is clear anatomical evidence that the diameter of the pupil shows opposing responses to local release of NAd and ACh: the pupil dilates following NAd released by the LC via the superior cervical ganglion (SCG; Fig.2A red) [18], whereas following ACh release the pupil constricts through the LDT driving the Edinger–Westphal nucleus (EW) via the ciliary ganglion (CG; peripheral parasympathetic ganglion; Fig. 2A blue) [21–23]. Furthermore these opposing systems are further enhanced by a mutual inhibition, whereby the LC inhibits the EW through $\alpha$2 receptors and the PPN inhibits the LC via cholinergic activation of LC GABAergic cells [18,21]. Given this clear anatomical dichotomy, it is thus surprising that levels of both NAd and ACh have been shown to covary with pupil dilation – specifically, a genetically encoded calcium indicator imaging of LC ($\alpha$1R) and nbM (M1R) cortical projections revealed a lagged response in which LC activity precedes both nbM activity and pupil dilation (Fig. 2b reproduced from [14]). The fact that these empirical results contrast with the underlying neuroanatomy makes it difficult use pupil diameter as an indirect readout of the adrenergic and cholinergic neuromodulatory system.

To resolve this conflict, we built a model incorporating the known directional interactions between the key hubs of the noradrenergic and cholinergic circuits controlling pupil diameter, along with the various structures that provide anatomical constraints (see methods). We drove the model with oscillatory phasic bursts from the LC that are known to activate the low-affinity $\alpha1$ adrenoreceptors [24,25], as these receptors are the targets of optogenetic imaging [14] and are hypothesized to play a crucial role in flexible behaviour [26]. By incorporating time-delays between the adrenergic and cholinergic systems, we were able to recreate the observed successive activation of LC, nbM (~0.5s delay), and pupil dilation (~1s delay; Fig. 2C). The model was also constrained to produce a pupillary oscillation that matched the natural frequency, ~0.2 Hz observed in the pupillary hippus [27]. Our model reproduced the tight relationship between LC activity and the first derivative of the pupil size, with a matched cross-correlation decay profile (Fig. 2D experimental, Fig. 2E model). Based on this lagged response of the model and the opposing anatomical roles of the LC and nbM in controlling the pupil, we predicted a temporal relationship between the relative balance of adrenergic and cholinergic tone and pupil fluctuations.

To test this hypothesis, we calculated the difference between the normalised activity of the LC and the nbM within the model, and found that the acceleration of the pupil size $d^2p/dt^2$ (Fig. 2F top black) was strongly positively correlated to the balance between systemic adrenergic and cholinergic tone (LC − nbM; Fig. 2F top purple) in the cerebral cortex (Fig. 2G top; red line indicates p < 0.05 95% standard-error of the mean (s.e.m)). When the normalised activity of the LC was greater than the normalised activity of the nbM, the pupil was positively accelerating – i.e., rapidly changing from constricted to dilated (and *vice-versa*). Following from this result, we reanalysed the empirical recordings of [14] (Fig. 2B) and observed the same relationship in the pupil-phase matched, trial-averaged responses (Fig. 2F bottom). Furthermore, comparing the lagged cross-correlation between LC − nbM and pupil acceleration, both the model and empirical displayed the same zero-lag peaked correlation and decay rate (Fig. 2G bottom). Thus, using a biophysical model and 2-photon calcium imaging, we have demonstrated a robust relationship between the rate of change of pupil size tracking LC activity (recapitulating empirical findings of [14]; i.e., $\frac{dp}{dt} \propto$ LC; Fig. 2H), and the relative difference between LC and nbM activity tracking the acceleration of pupil size (i.e., $\frac{d^2p}{dt^2} \propto$ LC − nbM; Fig. 2H), that are likely coordinated by phasic LC bursts. This analysis thus confirms that fluctuations in pupil diameter can be used as an indirect yet sensitive readout of neuromodulatory tone in the brain.

**Electrophysiological Evidence for Neuromodulatory Mediated Adaptive Dynamics**

With a better appreciation of the relationship between pupillary dynamics and neuromodulatory tone, to test our original hypothesis, we next required access to LV_PN neuronal activity across dynamic neuromodulatory states. To this end, we re-analyzed *in vivo* spontaneous electrophysiological neural activity recorded from two awake mice (*Mus musculus*) [28] with simultaneous recordings of the pupil (Fig. 3B). The neural data were collected from eight invasive silicon neuropixels probes (Fig. 3A) [29], which provided high-resolution access to multiple cortical neurons in awake animals. To isolate bursting neurons in Layer V, we applied two specific criteria: first, we selectively analyzed neuronal units that were identified on channels referenced to Layer V of the Allen Mouse Common Coordinate Framework [30]; and second, we identified units that demonstrated both bursting and regular spiking, defined as satisfying $b_{ISI}/n_{ISI} > 5\%$ and $b_{ISI}/n_{ISI} < 95\%$ where $b_{ISI}$ is the number of ISI < 10ms and $n_{ISI}$ is the total number of ISI, i.e. the neuron displayed at least 5% bursting spikes and no more than 95% bursting spikes. This approach ensured that we identified cells within Layer V that demonstrated both bursting (i.e., short ISI) and 'regular' spiking (i.e., long ISI; Fig. 2C) modes consistent with the dual spiking modes of LV_PN (Fig. 3C) and inconsistent with purely regular spiking thin-tuft LV pyramidal neurons or fast-spiking interneurons. This procedure identified 308 Layer V bursting units across eight visual and motor-sensory regions. To support this claim, we further confirmed that the mean burst ratio of the isolated units ($\beta_{mouse} = \frac{b_{ISI}}{n_{ISI}} = 0.26 \pm 0.13\ s.d$) was consistent with observed bursting ratios of LV_PN [11].

With estimates of LV_PN burst-firing and neuromodulatory tone in place, we set out to test the hypothesis that NAd and ACh differentially affect network-level information processing dynamics [14]. We predicted that fluctuations in pupil diameter, $d^2p/dt^2$, should differentiate distinct signatures of network-level neural dynamics. To measure these network-level properties in spiking data, we calculated the population spike count,

$$\rho_t = \frac{1}{N} \sum_i \delta(t - t_i),$$

which represents the number of spikes at time $t_i$ across $N$ neurons calculated in 1ms windows [31]. Neuronal firing rates showed large, coordinated deflections at both the individual neuron and population firing scale (Fig. 3D).

Theories of neuromodulation hold that adrenergic and cholinergic tone can shift the neuronal system between flexible dynamics, in which neuronal spiking is highly variable in both space and time, and reliable dynamics, in which neuronal spiking patterns are more spatiotemporally constrained [14]. Thus, phasic increases in NAd (i.e., pupil dilation positive $d^2p/dt^2$) should coincide with large fluctuations in $\rho_t$, whereas phasic increases in ACh (i.e., pupil constriction negative $d^2p/dt^2$ should quench changes to $\rho_t$. To assess these dynamic changes, we utilized time-varying estimates of variance:

$$\text{Var}(\rho_t) = \langle \rho_t^2 \rangle - \langle \rho_t \rangle^2,$$

calculated in non-overlapping 100ms windows (Fig. 3D teal box) and $\langle X \rangle$ signifies the expected value of $X$. Our analysis confirmed our hypothesis: specifically, we observed robust positive (negative) relationships between pupillary dilation (constriction) acceleration and $\text{Var}(\rho_t)$ ($p < 7 \times 10^{-6}$; Fig. 3E error-bars s.e.m across 16 electrodes - 2 recordings of 8 electrodes). Thus, an increase in NAd (ACh) tone leads to the enhancing (quenching) of the spiking variability and coherence of LV$_{PN}$ in recordings from an awake, freely behaving mouse *in vivo*. In this way, our results demonstrate a direct relationship between adrenergic neuromodulation and flexible dynamics and cholinergic neuromodulation and reliable dynamics at the network scale.

As LV$_{PN}$ are numerous, large, and geometrically aligned, their pooled spiking activity is thought to significantly contribute to extracellular field potentials recorded during ECoG [32–38]. As such, we wondered whether the network-scale relationship between pupil diameter and neuronal variability would be present at a mesoscale. To test this hypothesis, we utilised 18 ECoG recordings (6 mice) captured simultaneously with pupil measurements illuminated via an infrared light (Fig. 3F). The infrared light allowed for pupil recordings, even when the mouse had closed eyes, such as across sleep stages, thus allowing us access to recordings from rapid-eye-movement (REM) sleep, non-REM sleep and wake. Crucially, REM is known to be a purely cholinergic state [21], centrally, and hence allowed us to further test our original hypothesis and model. Utilising the predefined epochs from [39], we found that the pupil was significantly constricted during REM (when compared to awake; Fig. 3G; REM purple vs NREM green and Wake orange; p<0.001), and ECoG signals displayed small variance (Fig. 3H; p<0.001). We then analysed pupil diameter within the wake conditions using the same time-varying variance measures as above (i.e., 100ms windows) and found consistent results: specifically, inferred cholinergic (adrenergic) regimes displayed quenched (variable) dynamics (Fig. 3I error bars s.e.m. across 18 recordings). Thus, differential dynamical modes mediated by neuromodulation are present at both the micro- and meso-scales.

**Neuromodulation Shapes Spontaneous Dynamics in Nonlinear Layer V Pyramidal Neurons**

To date, most theoretical work into neuromodulation of neuronal networks has hypothesized that signatures of complex, adaptive dynamics emerge from complex interactions between coalitions of relatively simple (i.e., linear) 'point' neurons. However, this viewpoint has not yet embraced the unique nonlinearity of LV$_{PN}$ within the mammalian brain. To discern the mechanistic origin of this phenomenon, we constructed a spatially embedded nonlinear neuronal network, in which simulated neurons consisted of two coupled compartments – a somatic and an apical dendritic compartment – designed to mimic the dual compartment coupling of LV$_{PN}$ (Fig. 4A; see Methods for full details) [9]. Typical neuronal modelling focuses on either the single neuron – at a multi-compartment resolution – or a network of neurons – each at a point-neuron resolution. By utilizing our novel dual-compartment network model, we created biologically plausible spike profiles while retaining computational efficiencies [11]. In this way, we were able to examine the network-level interactions of ensembles of thousands of LV$_{PN}$ under a precise combination of neuromodulatory control that would be impossible *in vivo*.

To systematically examine the effect of neuromodulation on LV$_{PN}$ dynamics, we simplified our model parameter space to investigate two independent parameters influenced by neuromodulation. First, we investigated the probability, $\beta$, that apical drive exceeds the apical-somatic electrotonic separation mediated by the HCN leak current $I_h$ (Fig. 4B). $\beta$ ranges from $\beta = 0$ where LV$_{PN}$ cells cannot burst to $\beta = 1$ where LV$_{PN}$ cells only burst. Second, we investigated the spatial influence of correlated apical drive, $\sigma$ (Fig. 4B). This parameter captures differential profiles of apical drive to the system, ranging from bursting LV$_{PN}$ that are spatially uncorrelated (i.e., bursting cells are unlikely to be adjacent; $\sigma = 1$) to strongly spatially correlated ($\sigma = N$). We stimulated each parameter combination with identical white-noise drive to somatic and apical compartments before $\sigma$ apical spatial smoothing and analyzed the emergent spiking dynamics. The combination of nonlinear neurons and diffuse coupling was sufficient to create substantial heterogeneity in the model's emergent spiking dynamics. We confirmed that an even mixture of spatially correlated bursting and regular spiking (Fig. 4C purple) was associated with an elevated, albeit low mean pairwise spike-count correlation ($|r_{SC}| < 0.07$), which is consistent with experimental predictions [40].

Based on their differential neuroanatomy, NAd and ACh were hypothesized to have divergent effects on $\beta$ and $\sigma$. Both neuromodulators increase $\beta$, albeit through different molecular pathways: NAd promotes LV$_{PN}$ bursting via the $\alpha$2A receptor-mediated closure of HCN-gated I$_h$ channels [6]; and ACh by depolarising M$_1$ receptors [7]. Despite this similarity, the two systems have divergent effects on $\sigma$ [8]: adrenergic projections are diffuse and cross-regional boundaries in the cerebral cortex [41], whereas cholinergic projections are typically more segregated and excite somatostatin-expressing inhibitory neurons that further refine decorrelated apical spiking (Fig. 1A) [42,43]. For these reasons, in the [$\beta, \sigma$] model parameter space, the effect of NAd is conceptualized as a right-upward trajectory (red arrows; Fig. 4B), whereas ACh is conceptualized as a left-upward trajectory (blue arrows; Fig. 4B). Note that other mechanisms (not considered here) may move the brain along distinct state-space trajectories, including NMDA receptor engagement, which increases apical excitability [44] (i.e., an upward trajectory), or non-specific thalamic activity [45], which increases diffuse apical input and excitability (i.e., a right-upward trajectory).

We found that our model could recapitulate the shifts in spatiotemporal variability of LV$_{PN}$ under both NAd and ACh neuromodulation. Importantly, unlike *in vivo* experiments – which are fundamentally dynamic – we were able to stabilise the model to a constant neuromodulatory tone and hence, calculate spontaneous spike variability on the spiking outputs of our model. We initialised our NAd and ACh trajectories at the empirical bursting regime ($\beta_{mouse} \sim 0.25$ observed in awake mice (Fig. 4B, purple line, dashed line s.d. error bars) and a balanced apical input regime ($\sigma = N/2$) and we extended these trajectories linearly to extremal $\sigma$: $\sigma = 0.25 \times N$ (ACh) and $\sigma = 0.75 \times N$ (NAd). We found evidence to support the hypothesis that NAd and ACh differentially affect the spontaneous dynamics of the cerebral cortex. Specifically, we demonstrated that NAd increases spatiotemporal variability (Fig. 4D, increasing red) by engaging spatially correlated ensembles of bursting LV$_{PN}$. In contrast, ACh was found to quench spatiotemporal variability (Fig. 4D increasing blue) by engaging spatially separated ensembles of bursting LV$_{PN}$ (error bars corresponds to 95% confidence intervals across 100 simulations). These modelling results provide robust confirmation of the empirical signatures of the systems-level control of the ascending arousal system on the primary output cell of the cerebral cortex.

Our results indicate that the opposing adaptive modes of NAd and ACh emerge from their respective anatomical projections – the globally controlling diffuse adrenergic neuromodulation promotes variability and flexibility, whereas the locally controlling targeted cholinergic neuromodulation supports selectivity and reliability. Thus, our findings emphasize the importance of studying nonlinearities in neuronal systems, as this phenomenon would not be present in coalitions of linear point-neurons [46]. Our results suggest that neuromodulation can flexibly shift the spiking dynamics of LV$_{PN}$ to utilize the beneficial signatures of complex adaptive dynamics.

**Adrenergic and Cholinergic Neuromodulation Mediates Distinct Functional Information Processing Modes**
The previous analysis was under spontaneous conditions; however, our previous work [1] has argued that adrenergic and cholinergic neuromodulation alters how the system responds to stimuli. Based on these studies, we further hypothesized that cholinergic and adrenergic neuromodulation should differentially augment the network's receptivity to incoming stimuli [8]: NAd should augment flexibility and variability (albeit nonlinearly) [3], whereas ACh should enhance reliability and selectivity [47], analogous to widening or focusing the width of the flashlight, respectively. A robust method for probing a system's information processing capacity is to investigate its transfer function (or input-output/gain curve), which maps a precise input to a characteristic output. In psychophysics, a typical transfer function fit to experimental data is the power-law function $F(S) \sim S^\delta$, known as Stevens law, where S is the stimulus intensity, $\delta$ the scaling 'Stevens' exponent, and F(S) is the neural response to the stimulus [48]. An efficient transfer function possesses $\delta < 1$, allowing a given range of input stimuli to be mapped onto a smaller output range. We calculated the transfer function as,

$$F(S) = \frac{1}{T}\sum_{t=1}^{T} \rho(S),$$

which relates the mean spike density response, F(S), to afferent Poisson spikes, $\rho$, with a mean-rate S randomly distributed across the network. A useful metric that can be calculated from the transfer function is the dynamic range,

$$\Delta_S = 10\log_{10}\left(\frac{S_{0.9}}{S_{0.1}}\right),$$

which represents the range of discriminable stimuli [49]. The range $[S_{0.1}, S_{0.9}]$ are inverted from the transfer function $[F_{0.1}, F_{0.9}]$ with $F_x = F_0 + x(F_\infty - F_0)$ where $F_\infty$ and $F_0$ represent the saturation and baseline response, respectively. Another useful measure is the trial-to-trial variability of the transfer function,

$$\Delta_F = \langle \text{Var}(10\log_{10} F(S)) \rangle,$$

representing the intrinsic reliability or variability in mapping a stimulus to output [50].

To test our hypothesis, we calculated the transfer function in three regions of parameter space, which were chosen to represent the low-arousal state (Fig. 4E purple; $\sigma = 39$, $\beta = 0.2$), as well as under either adrenergic (Fig. 4E red; $\sigma = 52.3, \beta = 0.44$) or cholinergic (Fig. 4E blue; Fig. 4A blue; $\sigma = 25.7 \beta = 0.44$) neuromodulation. We found the three transfer functions all followed a power-law between their baseline and saturation values with the same scaling exponent $\delta \sim 0.8$, suggesting they efficiently map a large stimuli range to a smaller output, and that the psychophysical-law is invariant to arousal state (i.e., equivalent differences in stimulus lead to a proportional change in perceived magnitude across arousal). It should be emphasized that these properties are emergent phenomena and have not been coded into the network. For example, the low neuromodulation regime (purple) possessed the largest dynamic range, $\Delta_S$, (Fig. 4F). Increasing NAd led to the largest trial-to-trial variability, $\Delta_F$, (Fig. 4F; red), which is consistent with the theory that NAd facilitates flexible behaviour (i.e., a diffuse flashlight beam; 2). In contrast, increasing ACh led to a reduction in variability (Fig. 4F blue), corresponding to an increase in stimuli specificity and reliability, consistent with the known enhancement of stimulus detectability and focus with the increased cholinergic tone, i.e., a focused flashlight beam [47].

Our findings thus present an optimal solution to the inefficiency of fixed neuronal dynamics: the low neuromodulatory brain is associated with optimal signal detection (sensitivity), and two highly conserved neuromodulatory axes either sharpen specificity (reliability; ACh) or widen the responsivity (flexibility; NAd) of the system.

**Strong phasic bursts of the LC that sequentially activate the nbM facilitate novel brain state transitions**

Our findings thus far have shown that adrenergic and cholinergic neuromodulation differentially alter the information processing modes of the brain in both evoked and spontaneous regimes. However, these systems do not operate independently, as strong phasic bursts of the LC can engage the nbM via low-affinity $\alpha 1$ adrenergic receptors (Fig. 2A). This interdependency begs the question: What benefit does this neuromodulatory coupling confer to the brain?

Recent work has begun to interpret neural dynamics through low-dimensional brain-state trajectories [51–53], where the brain-state is the instantaneous representation of the underlying neural activity. This brain state trajectory then flows atop the underlying dynamic energy landscape [4]. Through this lens, common brain-states correspond to deep 'low-energy' wells in which the trajectory can get trapped such as an attractor state, whereas rarer brain-states correspond to raised 'high-energy' hills trajectory can readily flow away from such as a repellor state. Recent fMRI studies have shown that neuromodulators alter the energy landscape of neural BOLD dynamics, both in resting and task-evoked states [4] – ACh deepens energy wells, whereas NAd flattens the energy landscape [4]. Based on previous theoretical [8,54] and empirical work [1], we hypothesized that the modulation of the energy landscape detected in BOLD might emerge at the neuronal network scale.

To test this hypothesis, we calculated the energy landscape of the LV$_{PN}$ model across the three regimes analysed above. To do this we first estimated the 'energy', $E_{\Delta\rho}$, of changes in population firing rate, $\Delta\rho$, defined as $E_{\Delta\rho} = -\ln P_{\Delta\rho}$ where $P_{\Delta\rho}$ is the estimated probability of observing $\Delta\rho$ (see Methods). This technique has been applied to fMRI BOLD [4], MEG [55], spiking dynamics of retinal neurons [56,57], as well as the statistics of natural scenes [58]. Here, the term 'energy' follows a statistical physics definition that relates energy to the underlying probability distribution, and not the metabolic definition (i.e., the energy used by the brain to maintain or change neural activity) – i.e., a statistically energetically expensive state (such as changing firing rates) may be a metabolically energetically favourable state (and *vice-versa*). In this framework, a common change in firing rate $\Delta\rho$, such as when nearby an attractor, corresponds to a relatively low energy transition (i.e., low $E_{\Delta\rho}$), whereas a surprising change in state, i.e., large deviations from an attractor, corresponds to a higher energy transition (i.e., high $E_{\Delta\rho}$).

To test the hypothesis that adrenergic and cholinergic neuromodulation differentially alter the topography of the energy landscape, we calculated $E_{\Delta\rho}$ across the baseline/low-arousal state ($E_A$; Fig 5A purple), adrenergic ($E_{NAd}$; Fig 5A red), and cholinergic ($E_{ACh}$; Fig 5A red) regimes analysed above. We found that adrenergic neuromodulation flattened the energy landscape relative to the low-arousal state ($E_{NAd} - E_A$; Fig. 5B), thus equally facilitating all brain-state transitions [26], whereas cholinergic neuromodulation lowered the energy of small population firing rate transitions and raised the energy of larger population firing rate transitions, and thus emphasised the energetic separation between largely differing states, relative to the unaroused baseline regime ($E_{ACh} - E_A$; Fig. 5C).

While these results provided a clear mapping between energy landscape dynamics and neuromodulatory tone, the anatomy of the arousal system suggests that cholinergic hubs (such as the nbM) are recruited via $\alpha$1 adrenoceptors following phasic LC bursts. We naturally wondered what effect these connections might have on evolving neural dynamics. Upon closer inspection of the energy landscape diagrams (Fig. 5A-C), we found that the flattening of the energy landscape following LC activation will allow the brain to transition between different attractors (Fig. 5D; conceptually shown as wells in a 2D diagram where the brain state is represented by a ball and the arrows represent movement of this trajectory) whereas nbM activation will deepen the immediate attractor (Fig. 5E) by quenching large state-transitions, which is aligned with the role of ACh in focussed attention [2]. However, a flattened energy landscape is an inherently risky regime as the brain-state may randomly fluctuate to an undesired attractor. Thus, a natural solution to this risky regime is to immediately deepen the local energy landscape once the desired brain-state has been reached (Fig. 5F). This solution is exactly what occurs within the brain following phasic LC bursts (strongly flattening the energy landscape) that recruit the nbM (deepening the energy landscape) via $\alpha$1 adrenoceptors. In other words, the synergistic combination of LC mediated landscape flattening and cholinergic deepening can allow the brain to shift its neural dynamics to a desired new attractor (Fig. 5F bottom) that may have previously been difficult to reach (Fig. 5F top).

**Discussion**

In this study, we used a combination of empirical and modelling studies to demonstrate that adrenergic and cholinergic modulation adaptively facilitate flexible and reliable information processing modes in the brain, respectively. Specifically, we demonstrated a biological implementation for controlling different information processing modes in the brain by modulating the systems-level burst-firing dynamics of LV$_{PN}$ using the adrenergic and cholinergic arms of the ascending arousal system. Importantly, although both adrenergic and cholinergic inputs to LV$_{PN}$ increase bursting likelihood, the differential anatomical projections of the two systems are such that they can adaptively mediate distinct information processing modes, both during evoked and spontaneous conditions. Further, we found that the synergistic combination of NAd and ACh allowed neural dynamics to transition to otherwise difficult to reach novel brain-states. Our results also help to reduce confusion surrounding how adrenergic and cholinergic neuromodulation can be detected noninvasively through dynamic fluctuations in pupil diameter. Together, these findings help to refine our understanding of the dynamic neurobiological mechanisms responsible for adaptive behaviour.

Through our findings, we can re-interpret a number of canonical results from experimental neuroscience. The relationship between NAd tone and both cognitive function [3] and adaptability [26] can be reframed as arising from the augmentation of inherent nonlinearities within populations of pyramidal cells within the cerebral cortex. Specifically, diffuse adrenergic projections promoted spatiotemporal neuronal variability and flattened the energy landscape, which may facilitate adaptability and flexibility. Similarly, the association of ACh with heightened attentional focus [59] is likely mediated by the targeted cholinergic projections that quench spatiotemporal neuronal variability and lower the ongoing brain-states energy. This perspective helps to unify multiple unique conceptual vantage points and suggests novel empirical questions relating to other neuromodulatory systems and their precise anatomical targets in the brain.

We also demonstrated the presence of an interesting synergy between NAd and ACh. Strong, excitatory connections between the LC and nbM suggest that the two structures are inextricably linked to one another. These connections and our modelling results allow us a novel vantage point on classic results in the neuroscience literature, such as the association between LC activity and the so-called 'network reset' of evolving brain state activity [26]. Our results suggest the LC mediated 'network reset' is better described as a synergestic LC and nbM mediated effect, where LC activation allows the network state to be altered (via NAd flattening of the energy landscape) and the nbM enables the change state to be stabilised and implemented (via ACh deepening of the local energy landscape). Importantly, neither of these systems alone would be able to reset the network state.

Additionally, our results provide insight into the large variability in trial-to-trial responses to identical stimuli observed in animal recordings [60], whereby the slow-drift in response variability likely corresponds to fluctuations in neuromodulatory tone that ultimately change systems-level neural gain [1]. In particular, experiments have demonstrated quenched variability at the onset of stimulus trials [61], which in our model associates with an increase in cholinergic tone. We predict that other arms of the ascending arousal system, such as the dopaminergic, serotonergic, and histaminergic systems (to name a few), will play similar roles to controlling information processing modes, albeit constrained by the unique circuits that these systems innervate.

The ability to use pupillary dynamics as a non-invasive measure of different aspects of the neuromodulatory system offers various unique opportunities for systems neuroscience. For instance, by tracking both the raw pupil diameter as well as its first and second derivative, different behavioural states can now be linked to the balance between NAd and ACh, rather than either neurochemical alone. Confidence in this model can also be augmented by comparing pupillometry traces with direct recordings of the neuromodulatory system in the brain, perhaps using new sensitive markers of ACh [62] and NAd [63] in animals, preferably while they are awake and performing ethologically-relevant behaviours. Assuming these experiments confirm the predictions of our models, fluctuations in pupil diameter can then be used to create links between brain state recordings in different species, with the known caveat that there are species-specific idiosyncrasies relating to the pupil [17], in part due to differences in the propensity for nocturnal vs. diurnal behaviour.

Overall, our results suggest that a combination of both local (nonlinear apical-somatic mediated dynamics in $LV_{PN}$) and global (neuromodulatory control) control over neural dynamics is an effective means for efficiently modulating adaptive behaviour. Indeed, a number of critical brain processes, such as conscious perception [5,13] and the integration of feedforward and feedback signals in the cerebral cortex [11,64] have been linked to burst-firing in $LV_{PN}$, and also to the effects of neuromodulatory neurochemicals [7,65,66]. For this reason alone, understanding how neuromodulation alters $LV_{PN}$ neuronal dynamics is essential for improving our understanding of the brain, both in healthy and diseased states.

**Materials and Methods**

There are three sections to our methodological approach: 1) Experimental details describing the three mouse datasets analysed in the project; and 2) details of the biophysical pupil and Layer V pyramidal neuron modelling; and 3) the analytical techniques applied to the experimental and simulated data.

**Contact for reagent and resource sharing**

Please contact Brandon Munn (brandon.munn@sydney.edu.au) for resource sharing.

1. **Experimental details**

We analysed three open murine datasets throughout the manuscript, which will be described below in order of usage.

**Simultaneous adrenergic or cholinergic two-photon mouse and pupillary recording**

Experimental data presented in (Fig. 2c & Fig. 2e [14]) was reanalysed. The full description of the experiment can be found in [14]. Briefly, 21 mice underwent simultaneous two-photon imaging of either cholinergic (M1) or adrenergic ($\alpha$1a) axons expressing GCaMP6s within primary visual cortex and pupillary recordings. Calcium traces and pupil recordings were filtered between 0.1Hz to 10Hz. Pupil recordings were Hilbert Transformed and all periods of pupil constriction/dilation exceeding 1s were trial-averaged and analysed as presented in Fig. 2C. For our analysis into the relative balance between adrenergic and cholinergic (LC-nbM) activity we compared the relative difference between the standardised levels of the phase-aligned curves presented in Fig. 2C. Standardisation was calculated to allow a comparison between the trial-averaged responses as adrenergic/cholinergic recordings were not simultaneous. Simultaneous recordings to confirm our analysis would be well suited to be studied in future studies.

**Mouse Neuropixels recordings - Spike Sorting and Pupil analysis**

We analyzed freely available eight-probe Neuropixels recordings in three mice undergoing spontaneous conditions [28]. In each of the three recordings the eight-probe Neuropixels were placed in varying locations of the mouse brain which were conducted according to the UK Animals Scientific Procedures Act (1986) at University College London. The full protocol and experimental details can be found in [28]. Here, we briefly describe the aspects of the protocols that are relevant for our analysis.

The spiking data consisted of pre-processed units identified using a modified Kilosort algorithm (code available at www.github.com/MouseLand/Kilosort2) along with the height of each unit on the Neuropixel probe. The height of each unit was given in microns in the Allen CCF framework [30] which allowed cortical layers to be identified. Layer V units were defined as units within Layer V that are not adjacent to the boundary i.e., the next electrode is located in Layer IV or Layer VI. Layer V units were then further split depending on the spiking dynamics, we reasoned that LV$_{PN}$ should display a bimodal ISI distribution, defined as satisfying $b_{ISI}/n_{ISI} >$ 5% and $b_{ISI}/n_{ISI} < 95\%$ where $b_{isi}$ is the number of ISI<10ms and $n_{isi}$ is the total number of ISI (i.e., burst and regular spiking activity are present). This procedure identified 308 Layer V bursting units across visual, sensorimotor, and retrosplenial cortical areas.

Along with the electrophysiological recordings the mouses right eye was simultaneously recorded at 100 Hz (available at https://figshare.com/articles/Behavioral_videos_for_8-Neuropixels_recordings_from_Stringer_et_al_2019_Science/8378360). We analysed the pupillometry using the program Facemap (www.github.com/MouseLand/FaceMap), which returned an estimate for the pupil area at each frame. Briefly, the program iteratively fits a two-dimensional Gaussian centred on the darkest pixel of each frame. After calculating the approximate pupil area, the final pupil signal was band-passed filtered between 0.025 and 3 Hz, and large deviations from this signal were removed and cubically re-interpolated (typically blinks or noise) from the original signal before and this cleaned signal was band-passed filtered between 0.025 and 3 Hz. This final step was necessary to avoid large deviations when analysing the derivative of the signal.

**Mouse ECoG and Infra-red Pupillometry**

Mouse ECoG and Pupillometry analysed recordings made from the Huber Lab presented in [39] where full experimental details can be found. Briefly, simultaneous infrared back-illumination pupillometry and M1 ECoG were recorded from six mice (18 recordings). As the mice were habituated to the head-fixed position they were allowed to sleep and the infra-red backlit allowed pupil size to be recorded with eyes closed. The sleep-staging performed by [39] using both ECoG and EMG activity was used to analysis pupil size and ECoG variance across varying arousal stages (REM, NREM, and Wake).

**2. Biophysical modelling**

**Adrenergic and Cholinergic Pupil Model**

The results presented in Fig. 2 were obtained via numerical simulation using a biophysical model of the regions identified in Fig. 2A as relevant for relating adrenergic activity (via the LC) and cholinergic activity (via the LDT/PPN/nbM) with pupil fluctuations. This model is by no means a complete model of pupil fluctuations which are controlled by many other regions. The model contained the following regions the LC, LDT/PPN, nbM, EW, SCG, CG, and the pupil diameter, $p$, which were simulated using the following equations:

$$LC(t) = lc_I(t) + C_{LDTPPN,LC} LDTPPN(t - \tau_{LDTPPN,LC}),$$

$$LDTPPN(t) = ldtppn_I(t) + C_{LC,LDTPPN} LC(t - \tau_{LC,LDTPPN}),$$

$$nbM(t) = nbM_I(t) + C_{LC,nbM} LC(t - \tau_{LC,nbM}),$$

$$EW(t) = EW_I(t) + C_{LC,EW} LC(t - \tau_{LC,EW}) + C_{LDTPPN,EW} LDTPPN(t - \tau_{LDTPPN,EW}),$$

$$CG(t) = C_{EW,CG} EW(t - \tau_{EW,CG}),$$

$$SCG(t) = C_{LC,SCG} LC(t - \tau_{LC,SCG}), \&$$

$$p(t) = p(t-1) + C_{SCG,p} SCG(t - \tau_{SCG,p}) - C_{CG,p} CG(t - \tau_{CG,p}) - p_b,$$

where $C_{A,B}$ and $\tau_{A,B}$ represents the connection weighting and time-delay from region $A$ to $B$, respectively. $A_I(t)$ represents the external input into region $A$ at time $t$. $p_b$ is a baseline normalisation constant $p_b = \langle C_{SCG,p} SCG(t - \tau_{SCG,p}) - C_{CG,p} CG(t - \tau_{CG,p}) \rangle$. For the simulations presented in Fig. 2 we drove the system only via the LC with an oscillatory drive of $f_{LC} = 0.25$ Hz bound between 0 and 1, i.e. $lc_I = 0.5 \times (-\cos(2\pi f_{LC}) + 1)$ and $ldtppn_I$ & $nbM_I = 0$. Simulations were calculated with a timestep of 0.2s or 50Hz sampling. Parameters were fit to match the empirically observed oscillation frequency, lagged response, and correlation between LC, nbM and $p$ and LC and $\frac{dp}{dt}$. Further justification behind specific connections can be found in the Results section.

The fitted parameters were given by:

| | |
|---|---|
| $C_{LDTPPN,LC}$ | 0.5 |
| $\tau_{LDTPPN,LC}$ | 0.2 s |
| $C_{LC,LDTPPN}$ | 1 |
| $\tau_{LC,LDTPPN}$ | 0.6 s |
| $C_{LC,nbM}$ | 1 |
| $\tau_{LC,nbM}$ | 0.6 s |
| $C_{LC,EW}$ | 1 |
| $\tau_{LC,EW}$ | 0.2 s |
| $C_{LDTPPN,EW}$ | 0.5 |
| $\tau_{LDTPPN,EW}$ | 0 s |
| $C_{EW,CG}$ | 1 |
| $\tau_{EW,CG}$ | 0 s |
| $C_{LC,SCG}$ | 1 |
| $\tau_{LC,SCG}$ | 0.2 s |
| $C_{SCG,p}$ | 0.5 |
| $\tau_{SCG,p}$ | 0.2 s |
| $C_{CG,p}$ | 0.5 |
| $\tau_{CG,p}$ | 0.2 s |

**Dual-compartment Layer V Pyramidal Neuron model**

The results in Fig. 4 and Fig. 5 in the main text were obtained via numerical simulation using a phenomenological quadratic adaptive integrate and fire neuronal model proposed by Izhikevich [67], which is a canonical reduced form of Hodgkin-Huxley neuronal dynamics [67], conserving key aspects of their original dynamics (i.e., spike generation and bursting), but distils the full equations to a two-dimensional system of ODEs, with 4 dimensionless parameters that can be modified to recapitulate a range of spike-adaptation dynamics that have been observed experimentally. Within the model the somatic dendritic compartment determines the generation of the spike wave form and dynamics, and the apical dendritic compartment serves to shift the spike-adaptation of the somatic dynamics between a mode of regular spikes (no apical intervention) and one of bursting (apical intervention), dependant on the activity within apical dendritic compartment.

**Somatic dendritic compartment**

First, we define the dynamics of the somatic compartment which generates the spikes. The somatic dendritic compartment was modelled by the membrane equation,

$$\frac{dv}{dt} = h(0.04v^2 + 5v - u + I), \quad (1)$$

$$\frac{du}{dt} = h(a(b(v - v_r) - u)), \quad (2)$$

with the after-spike resetting given by

$$\text{if } v \geq 30 \text{, then } \begin{cases} v \leftarrow c(t) \\ u \leftarrow u + d(t) \end{cases}, \quad (3)$$

where the differential equations are in a dimensional form, such that the membrane potential, $v$, is in millivolts (mV), $t$ is time in milliseconds (ms), $v_r$ is the resting potential in millivolts (mV), and $u$ is the recovery variable, defined as the difference of all inward and outward voltage-gated currents (this emulates the activation (inactivation) of potassium (sodium) ionic currents). $I$ is the current into the somatic dendrites from all sources in picoamperes (pA), and $h$ is the integration step, which was set at 0.5 ms so as to obtain optimal accuracy when simulating Izhikevich neurons. All differential equations in this work were solved using a 2nd-order Runge-Kutta (RK-2) numerical integration method, and all analysis was computed on spike-times rounded to the nearest millisecond.

The parameter $a$ (ms$^{-1}$) represents the time constant of the spike adaptation current and is set as $a = 0.02$ ms$^{-1}$. The parameter $b$ (nS) describes the sensitivity of the adaptation current to subthreshold fluctuations of the membrane potential ($v$) and is set as $b = 0.2$ nS. These parameters and the constants in Eq.1&2 are reductions from Izhikevich to match the spike width to experimentally observed durations (see [67] for further details).

The parameters $c$ and $d$ represent the after-spike reset of $v$ and $u$, controlling the voltage reset to model the effect of fast high-threshold K+ conductances and the slow high-threshold Na+ and K+ conductances activated during the spike similarly modulating spike-adaptation as $a$, respectively. The parameters $c$ and $d$ are time-varying and are modified by the apical compartment, as detailed below.

In this paper, we studied a highly recurrent network of LV$_{PN}$, along with the role of spatially dependent apical input. Thus, we construct a network consisting of $N^2 = 70 * 70 = 4900$ neurons with toroidal topology (periodic boundary conditions). Afferent connections are made with adjacent neurons falling within a somatic dendritic tree of radius of 200 $\mu$m [68]. Total synaptic currents, $I$, (pA; Eq. 1) into the somatic dendrites is prescribed by

$$I = I_{ext} + s, \quad (4)$$

where $I_{ext}$ represents the external, non-specific input onto the LV$_{PN}$ from lower-cortical feed-forward and subcortical structures. This parameter was modelled as white noise ($\mu_{I_{ext}} = 0$ pA, $\sigma_{I_{ext}} = 5$pA) to induce spontaneous activity. For a given neuron $i$, $s_i$ represents the synaptic input from all afferent neurons, while additionally incorporating inhibitory and excitatory neurons. We utilised a homogenous network connectivity utilising a sum of two exponentials (i.e., Mexican-hat coupling) to model the local excitation and lateral inhibition effects [69]. The total synaptic current into a neuron, $i$, is then given by:

$$s_i(t) = \sum_j \sum_k w_{ij}\, \delta(t - t_j^k), \quad (5)$$

where $\delta$ is the Kronecker delta function and spikes at time $t^k$ from all afferent neurons, $j$, are scaled by a synaptic coupling weight, $w_{ij}$, and summed. The synaptic coupling strength follows a difference of Gaussians or 'Mexican-hat' function [69] given by

$$w_{ij} = \begin{cases} 0 \text{ if } d_{ij} > d_{max} \text{ or } i = j \\ C_E e^{-\frac{d_{ij}^2}{d_E}} + C_I e^{-\frac{d_{ij}^2}{d_I}} \text{ if } 0 < d_{ij} < d_{max} \end{cases}, \quad (6)$$

where $d_{ij}$ is the Euclidean distance between neuron $i$ and $j$, $C_E$ and $C_I$ are the excitatory and inhibitory coupling constants, and $d_E$ and $d_I$ are the excitatory and inhibitory coupling ranges, respectively. The coupling parameters were set in our model such that the model was poised on the border between "dying" and "run-away" activity, which ensures that the excitatory and inhibitory strengths are consistent with spike rates of spontaneous cortical activity in humans (i.e., ~4 Hz), furthermore they were normalized to the model size to ensure network rescaling preserved approximate spiking dynamics. The coupling parameters utilized in our simulations were $C_E = 180/N$, $C_I = C_E/2$, $d_E = N(6)$, $d_I = 2N$, $d_{max} = 2.5N$. Finally, the parameters used in the model are set such that the network is balanced, defined as $\sum w_{ij} = 0$, i.e., the net synaptic coupling into each neuron is zero.

**Apical compartment**

The somatic dendritic compartment of each neuron was coupled to its corresponding apical dendritic compartment, whose increased activity could thus transition the neurons somatic dendritic compartment from a regular spiking mode to a burst spiking mode, in a process known as 'apical amplification' [44]. The presence of coincident apical drive that exceeded the apical-somatic electrotonic separation within a 30ms window [64] caused the apical compartment to switch the somatic spiking properties from empirically observed LV$_{PN}$ regular spiking ($c(t) = -65, d(t) = 8$) to a burst firing behaviour ($c(t) = -55, d(t) = 4$) [70].

**The spatial profile of apical input, $\sigma$**

The apical compartment input, $\alpha(t)$, was modelled as receiving normalized white noise drive, which was either targeted to an individual neuron or shared diffusely between neighbouring neurons depending on a diffusivity parameter, $\sigma$. The diffusivity parameter ranges from $\sigma = 1$, targeted apical input, to $\sigma = N$, diffuse apical input. The apical input was then calculated by generating a unique white-noise drive for each apical node and then spatially convolving this with a Gaussian kernel and integrating over the previous 30ms as follows:

$$\alpha_i(t) = \sum_{t'=t-30}^{t} G_\sigma(d_{ij}) * I_{ap}(t'),$$

$$\alpha_i(t) = \sum_{t'=t-30}^{t} \sum_j G_\sigma(d_{ij}) I_{ap}(t'), \quad (7)$$

where $I_{ap}(t)$ is the current into the apical dendrites at time $t$ given by dimensionless normalized white-noise, $d_{ij}$ is Euclidean distance between nodes $i$ and $j$, and $G_\sigma(d_{ij})$ is a Gaussian kernel, $G_\sigma(d_{ij}) = \frac{1}{2\pi\sigma^2} e^{-\frac{|d_{ij}|^2}{2\sigma^2}}$. Each apical current, $\alpha_i(t)$, is then normalized to ensure the convolution with the Gaussian kernel preserves the scale of the apical input across differing model parameters.

**Burst parameter, $\beta$**

The apical input can transition each LV$_{PN}$ from a regular spiking mode to a burst spiking mode, depending on whether the input exceeds the hyperpolarization-activated cyclic nucleotide–gated (HCN) channel-mediated leakage current, $I_h$, which controls the burst probability $\beta$ [66,71]. The coupling to the somatic dendrite, via the spiking variables $c(t)$ and $d(t)$ in Eq. 3, is given by

$$c(t) = -65 + 10H(\alpha(t) - I_h), \& \quad (8)$$
$$d(t) = 8 - 4H(\alpha(t) - I_h), \quad (9)$$

with $H$ as the Heaviside step-function. This results in two conditions: if the apical current does not exceed the HCN channel-mediated leakage current, then $c = -65$ mV and $d = 8$, and the neuron recapitulates regular spiking dynamics, such that when driven with constant (DC) current the neuron responds with a short inter-spike interval (ISI) which gradually increases with current amplitude; in contrast, if the apical current exceeds the HCN channel mediated leakage current then, $c = -55$ mV and $d = 4$, which recapitulates intrinsically bursting spike dynamics, such that when driven with constant (DC) current, the neuron responds with bursting, followed by repetitive short ISI spikes. The two regimens of spiking dynamics can be observed in Fig. 1b in the main text. The parameters were chosen based on original research that fit the spike profiles of regular spiking and bursting LV$_{PN}$ [67].

**Exploring the model state-space according to $\beta$ and $\sigma$**

We oriented our model to explore the two key parameters corresponding to burst probability, $\beta$ and the extent of spatially correlated apical input, $\sigma$ which are both differentially modified by adrenergic and cholinergic neuromodulation. Since the model is driven by a pre-defined white-noise input into the somatic compartment, $I_{ext}$, and apical dendritic compartments $I_{ap}$ (prior to convolution with gaussian spatial kernel) we can effectively isolate the effect of neuromodulation via its modification of $\beta$ and $\sigma$. In the manuscript, we present results for an HCN channel mediated leakage current ranging from $I_h = -3$ ($\beta = 0$) to $I_h = 3$ ($\beta = 1$) in 40 linear steps, and $\sigma$ ranging from $\sigma = 1$ (targeted apical input) to $\sigma = 70$ (diffuse apical input as if there was a common apical dendrite) in 42 linear steps. Thus, we ran $40 \times 42 = 1680$ simulations, with identical temporal drive and apical input, prior to Gaussian convolution. We ran each simulation for 35 s, in time-steps of $\Delta t = 0.1$ ms and discarded the initial 15s of simulation to avoid transient dynamics induced by initial conditions.

## 3. Analysis methods

**Inter-spike interval**

Inter-spike intervals (ISI), defined as the time interval between successive spikes in a spike train, were calculated for each neuron. Given $J$ spikes let $t_i$ be the occurrence time of the $i$th spike. The ISI sequence is:

$$\text{ISI} = \{t_2 - t_1, t_3 - t_2, \ldots, t_J - t_{J-1}\}.$$

**Spike count**

To calculate spike-counts, we followed the approach described by [40]. First, time was divided into $dt = 1$ms bins, and a binary spike train, $\rho_i$ was created for each neuron, $i$, equal to 1 if there was a spike in $(t, t + dt)$ and 0 otherwise,

$$\rho_i = \sum_i \delta(t - t_i).$$

The population spike count, $\rho$, was then calculated as the cumulative spikes at a given timestep within the network model or recorded on a similar neuropixels probe. We calculated the population spike count,

$$\rho = \frac{1}{N} \sum_i \delta(t - t_i),$$

which represents the number of spikes at time $t_i$ across $N$ neurons calculated in 1ms windows [31]

**Time-varying variance analysis**

Throughout the manuscript we calculated the variance of time-varying population measures, this was calculated as,

$$\text{Var}(X) = \langle X^2 \rangle - \langle X \rangle^2$$

where $\langle X \rangle$ signifies the expected value of the signal $X$. The results presented in Fig. 3E&I display the variance of the Layer V multiunit or ECoG signal, respectively calculated within 100 ms windows. The time-varying variance was then normalised and binned between -3 to 3 in steps of 0.25, from which the mean and 95% standard error of the mean within each bin is presented.

**Dynamic range**

To probe the information processing properties of the brain given stimuli, we calculated the dynamic range, $\Delta$, from the range of discernible responses to the range of stimulatory input, also known as the transfer function or gain curve. We calculated the transfer function (i.e., response function or gain-curve), $F$, as

$$F = \sum_{t=1}^{T} \rho(t), \tag{25}$$

that is the spiking activity generated over $T$, where $T = 500$ ms (results are robust for varying $T = 100$ ms to $T = 2$ s), in response to a stimulus of strength $S$, ranging from $S = 10^{-5}$ to $S = 10^{0.5}$ ms$^{-1}$, where the stimulus is modelled as afferent spikes generated as a Poisson process to each neuron at a stimulus rate $S$. Finally, $F$ was averaged across 20 trials for each stimulus intensity. After calculating the average elicited response, F, the dynamic range, $\Delta = 10 \log_{10} \frac{S_{0.9}}{S_{0.1}}$, was calculated as the stimulus range (in dB) where variations in $S$ can be robustly coded by variations in $F$, after discarding elicited responses that are too small to distinguish from baseline, $F_0$, or network saturation, $F_{max}$ [49]. The stimulus range $[S_{0.1}, S_{0.9}]$ is calculated from the elicited response range $[F_{0.1}, F_{0.9}]$, where $F_x = F_0 + x(F_{max} - F_0)$ which is the standard range reported [49,50,72,73]. Finally, the trial-to-trial variability was calculated as the mean variance of the log-transformed gain-curve across 20 trials, where we log-transformed the data to ensure equal weighting to the variability of the low stimuli strengths.

**Fitting a power-law to the transfer function**

When plotted on a logarithmic axis, we found the three brain-states follow a power-law relationship, $F \propto S^\alpha$. We fitted a slope to the power-lag segments of each gain curve using a Levenberg-Marquardt nonlinear least-squares fitting algorithm and found the slope is consistent across each state $\alpha \sim 0.8$, a similar exponent observed across some psychophysical laws [48,74]. This result suggests both that the population is efficiently compressing a broader range of stimuli into a smaller range of response output, demonstrated by a slope exponent $\alpha < 1$, and that the psychophysical-law is invariant to brain-state changes (i.e., equivalent differences in stimulus lead to a proportional change in perceived magnitude across brain-states).

**Energy landscape analysis**

To quantify the change in Layer V neuronal activity following varying levels of adrenergic and cholinergic neuromodulation we calculated the population spiking activity squared displacement ($\Delta\rho$) and converted this to a statistical energy using an approach introduced in [4]. The population spiking activity squared displacement is a measure of the deviation in spiking activity across time. $\Delta\rho$ is calculated as the squared change of population spiking activity

$$\Delta\rho = |\boldsymbol{\rho}_{t_0+t} - \boldsymbol{\rho}_{t_0}|^2,$$

where the change in spike rate, $\boldsymbol{\rho}$, is calculated across different timelags $t$ (from 2ms to 5s in steps of 1ms), and $t_0$ are all timesteps from 1 to $T - \max(t)$. We are interested in the probability, $P_{\Delta\rho}$ that we will observe a given change in activity. We estimated the probability distribution function from all available $n$ datapoints $\Delta\rho$ – using a Gaussian kernel density estimation $P_{\Delta\rho} = \frac{1}{4n}\sum_{i=1}^{n} K\left(\frac{\Delta\rho}{4}\right)$, where $K(u) = \frac{1}{2\sqrt{\pi}} e^{-\frac{1}{2}u^2}$. As is typical in statistical mechanics the energy of a given state, $E_{\Delta\rho}$, and its probability are related $P_{\Delta\rho} = \frac{1}{Z} e^{-\frac{E_{\Delta\rho}}{T}}$, where $Z$ is the normalisation function and $T$ is a scaling factor equivalent to temperature in thermodynamics [56]. In our analysis $\sum_\sigma P_\sigma = 1 \to Z = 1$ by construction and we can set $T = 1$ for the observed data. Thus, the energy of deviation in spiking activity is then equal to the negative natural logarithm of the probability,

$$E_{\Delta\rho} = -\ln(P_{\Delta\rho}).$$

**Acknowledgments**

We would like to thank Matthew Larkum, Matthew McGinley, Michael Breakspear, Russell A. Poldrack, Christopher Whyte, Rick Shine, and Paul Martin for their constructive notes on the manuscript. Daniel Huber, Özge Yüzgeç, and the Huber Lab for sharing their pupil and ECoG recordings. Jacob Reimer for permission to reanalyse his published data. The Harris Lab and Janelia for their public repository of Eight-probe Neuropixels recordings in three mice. JMS was supported by an NHMRC Investigator grant (#1193857) and a University of Sydney Robinson Fellowship.

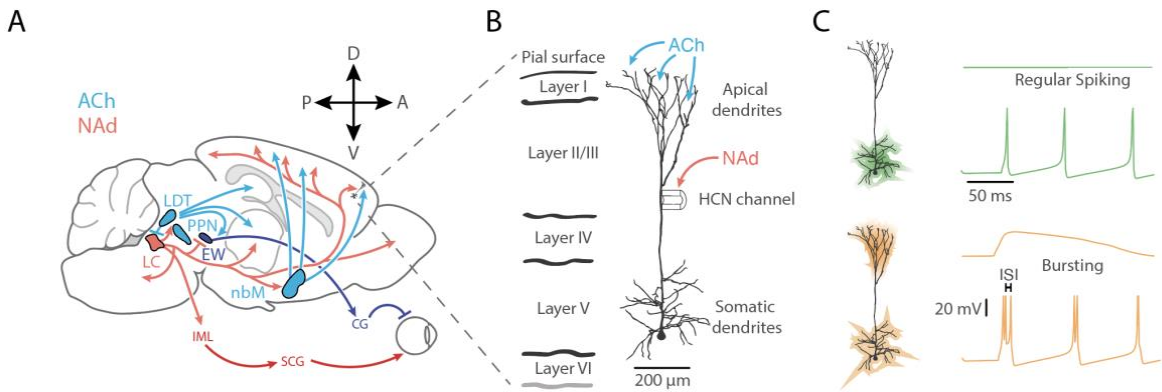

**Figure 1 – Adrenergic and Cholinergic neuromodulation differentially target nonlinear layer V pyramidal neurons.** A) The Locus Coeruleus (LC; rose) and nucleus basalis of Meynert (nbM; blue), which are predominantly responsible for cortical noradrenergic (NAd) and cholinergic (ACh) metabotropic neuromodulatory projections, contact layer V pyramidal neurons (LV$_{PN}$) in a diffuse and targeted manner, respectively. Within the brainstem, the pedunculopontine (PPN) and laterodorsal tegmentum (LDT) provide cholinergic tone to key subcortical regions. These systems intersect with pathways controlling pupil dilation (red) and constriction (navy) via the intermediolateral cell column (IML; within the spinal cord) then superior cervical sympathetic ganglion (SCG; within the sympathetic nervous system), and through the Edinger–Westphal nucleus (EW) via the ciliary ganglion (CG; peripheral parasympathetic ganglion) then the sphincter pupillary muscles, respectively. Saggital slice diagram of the mouse brain with coordinate directions dorsal (D), ventral (V), anterior (A), and posterior (P). B) Layer V pyramidal neurons (LV$_{PN}$) in the cerebral cortex span all cortical layers and consist of two dendritic compartments (apical and somatic dendrites) that are electrotonically separated by hyperpolarisation-activated cyclic nucleotide-gated (HCN) channels located along the apical trunk. C) LV$_{PN}$ are capable of nonlinear spiking dynamics, depending upon the stimulation location along the neuron's dendrites. The electrotonic separation is such that the neuron typically undergoes regular spiking (green) when driven by somatic drive; however, when the HCN-mediated electrotonic separation is exceeded within a short temporal window, the simultaneous apical and somatic drive can switch the neuron into a burst spiking mode (short inter-spike interval [ISI]; yellow).

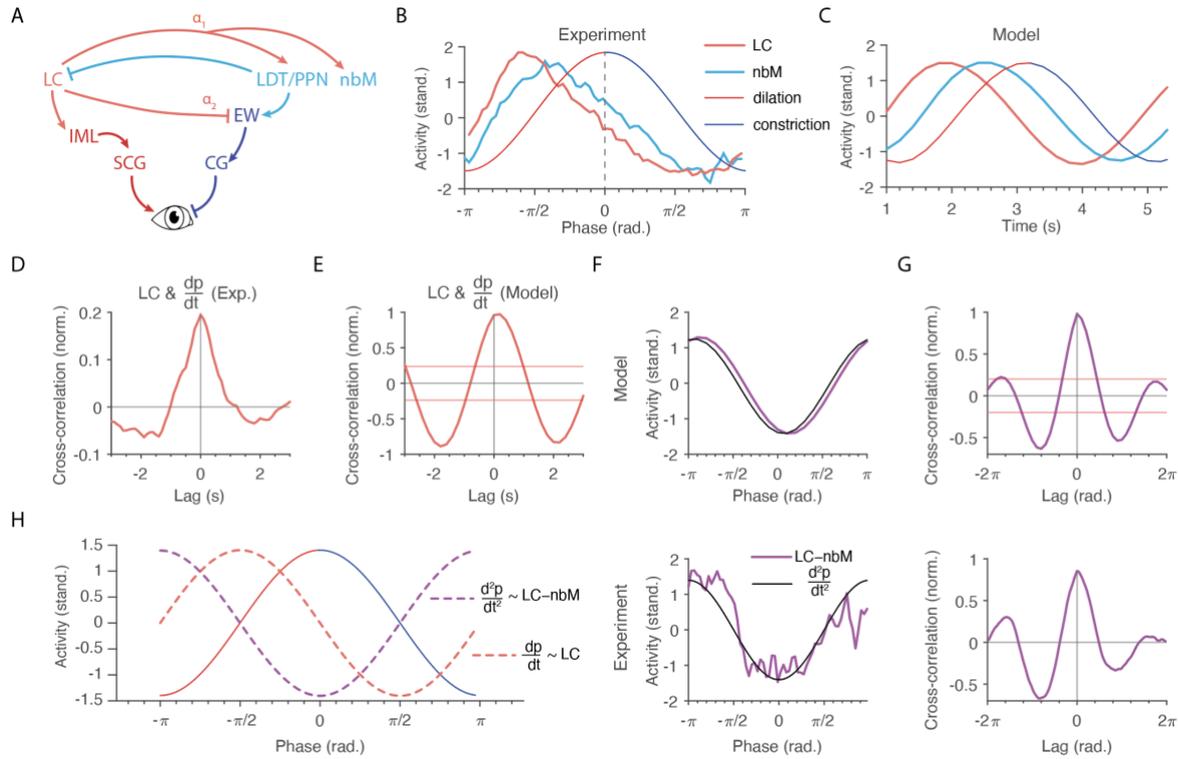

**Figure 2 - Pupil fluctuations track the balance between adrenergic and cholinergic neuromodulatory levels.** A) Anatomy of adrenergic and cholinergic neuromodulatory sites and their anatomical connections to pupil size. B) Experimental recordings of trial-averaged pupil phase relationship (dilation red, constriction navy) between LC (rose)and nbM (blue) efferent cortical connections show a lagged phasic relationship. Reproduced with permission from [13]. C) We created a biophysical model containing the regions outlined in (A) and matched dynamics to recreate the lagged relationship observed experimentally in (B). D) Experimentally, we observed a tight correlation between LC activity and the first derivative of the pupil size ($dp/dt$), which is recapitulated by the model (E). F) (Top) The model predicts a synchronous relationship between the relative balance of NAd and ACh (NAd – ACh) and the second derivative of the pupil size ($d^2p/dt^2$; acceleration). (bottom) This relationship is observed in the reanalysed difference between trial-averaged to pupil phase (NAd – ACh). G) The cross-correlation of pupil acceleration and the balance of NAd and ACh predicted by the model (top) matches the temporal lags observed experimentally (bottom). H) Analysis of the model and data suggest LC levels can be interpreted from the first derivative of pupil size (velocity, red dashed) and the balance between LC and nbM, corresponding to the relative NAd – ACh neuromodulatory tone can be inferred from the second derivative of the pupil size (acceleration, purple dashed).

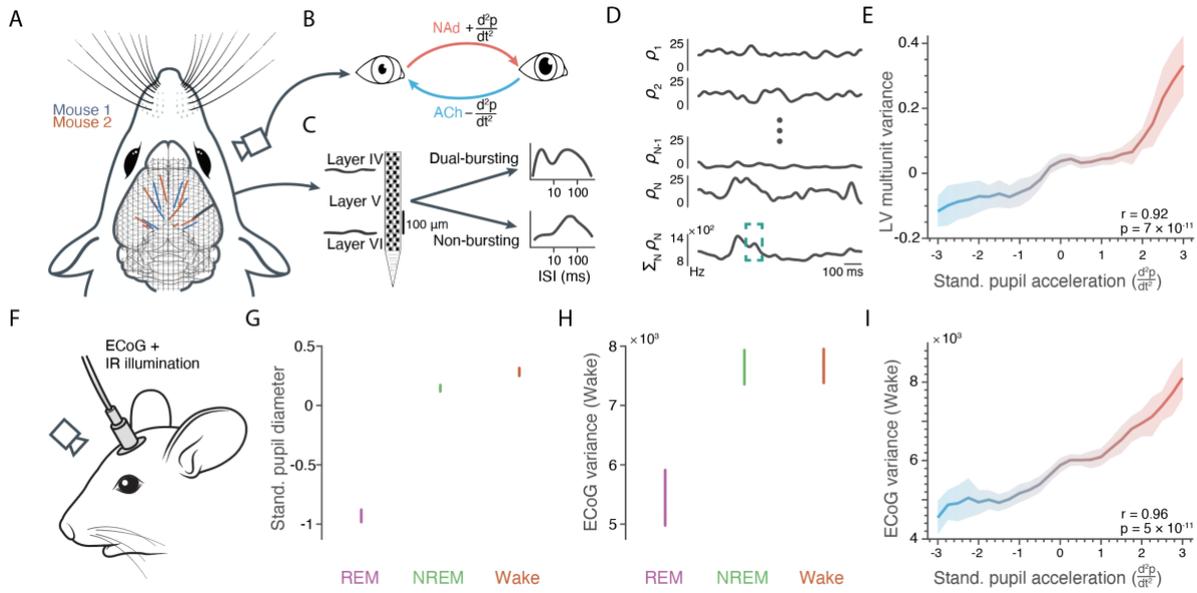

**Figure 3 – Electrophysiological Evidence for Neuromodulatory Mediated Adaptive Dynamics.** A) Under awake conditions, simultaneous pupillometry and spiking activity were recorded from 2 mice each using eight neuropixels distributed across the mouse cortex. B) pupillometry was analysed for the balance between adrenergic and cholinergic tone, a positive change in pupil acceleration (dilation) corresponds to an increase in NAd or decrease in ACh (adrenergic tone) and vice versa for a pupil deceleration (constriction; ACh >NAd). C) We isolated 308 units recorded off channels located within Layer V that possessed both bursting (isi<10 ms) and regular spiking. D) We calculated the firing rate, $\rho$, of each Layer V unit and quantified the dynamics of the system by the temporal variability of the pooled population firing activity (teal). E) To calculate these measures at the rate of pupil fluctuations, we calculated the instantaneous approximation of the spatiotemporal variability of population spiking that increased during adrenergic neuromodulation ($d^2p/dt^2 > 0$; i.e., pupil dilating) and was quenched during cholinergic neuromodulation ($d^2p/dt^2 < 0$; i.e., pupil constricting). F) ECoG and infra-red (IR) illumination allowed the simultaneous recordingof pupil and mesoscale electrophysiology. G) Standardised pupil diameter across the three arousal stages rapid-eye movement (REM; purple) and non-REM (NREM; green) sleep and Wake (orange). H) ECoG variance within the same defined epochs. G/H) errorbars represent 95% s.e.m. I) Time-varying eseiamtes of ECoG variance within the wake conditions recapitulates the neuronal spiking relationship of (E) that is increased spatiotemporal variability with adrenergic tone (assessed by pupil dilation $d^2p/dt^2 > 0$) and vice versa for inferred cholinergic tone (assessed by pupil constriction $d^2p/dt^2 > 0$).

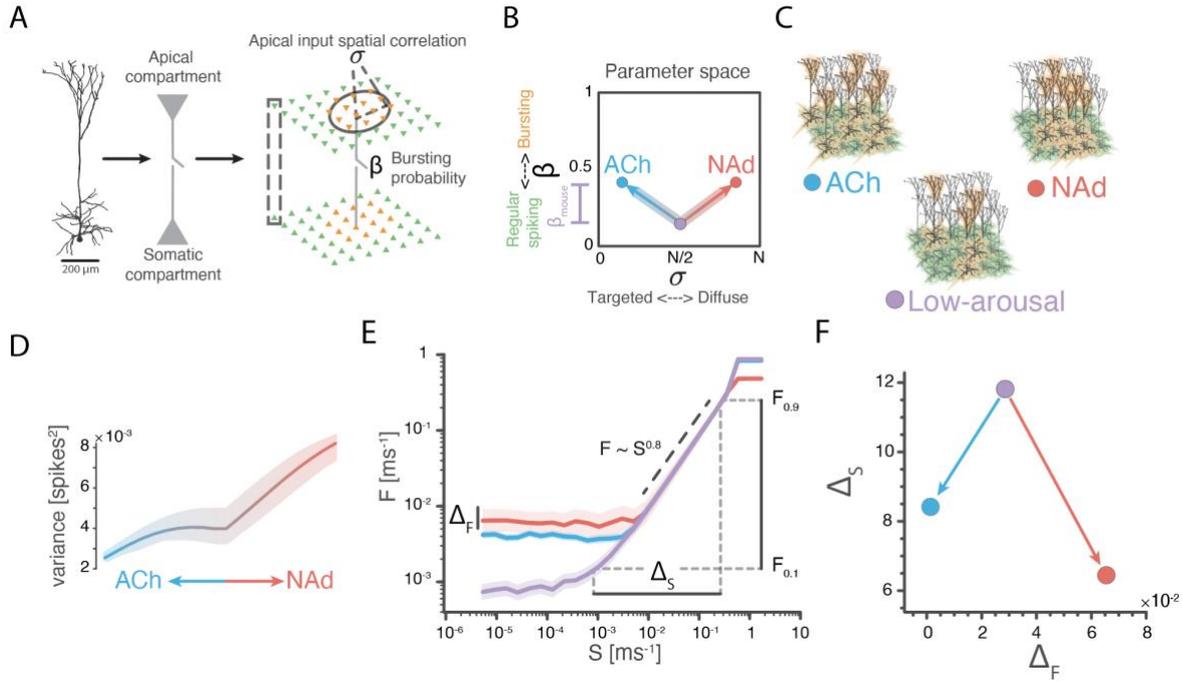

**Figure 4 – Adrenergic and Cholinergic Neuromodulation Mediates Distinct Functional Information Processing Modes.** A) We simulated a neuronal network of LV$_{PN}$ with apical and somatic compartments and isolated the effects of neuromodulation (red and blue alterations correspond to NAd and ACh, respectively) of the cells inherent nonlinearity by changing two parameters: the LV$_{PN}$ burst probability ($\beta$) and the apical input spatial correlation ($\sigma$). B) Spiking data were simulated across a [$\beta, \sigma$] model parameter space, within which we could track the trajectories of increasing neuromodulatory tone: NAd (red) and ACh (blue)both increase bursting probability; however, NAd increases the apical input spatial correlation due to the LC diffuse projections, whereas cholinergic projections are known to be more segregated and decrease the spatial correlation of inputs to apical dendrites (Fig. 1A). The neuromodulatory trajectories begin from the empirically observed mean burst probability of mouse Layer V $\beta_{mouse} \sim 0.25$ (error bar s.d.) and a balanced apical input regime ($\sigma = N/2$) and extend beyond the empirically observed regime to maximal apical input extremes ($\beta_{max} \sim 0.75$). C) Taken together the differential neuromodulatory projections and coupling to LV$_{PN}$ bursting means ACh will increase LV$_{PN}$ bursting in a targeted spatially uncorrelated manner (i.e., ungrouped) whereas NAD will increase LV$_{PN}$ bursting in a diffuse spatially correlated manner (i.e., grouped). D) Temporal spiking variability as quantified by the population spiking variability (mean and 95% confidence interval across 100 trials). E) We analysed the information processing of the neural system in the baseline neuromodulatory regime from Fig. 4B (purple) and two regions corresponding to either a cholinergic (blue) or adrenergic (red) phasic burst. The transfer functions between input stimuli intensity S and mean output F across repeated trials, reveals similarities and differences between the three regimes. The dynamic range, $\Delta_S$, and trial-to-trial variability, $\Delta_F$ were also calculated. F) the low arousal (purple) had the highest $\Delta_S$ (i.e., the highest sensitivity) and moderate $\Delta_F$, whereas ACh (blue) minimised $\Delta_F$ and thus increases specificity and signal reliability, whereas NAd (red) maximised $\Delta_F$ and hence promoted a more flexible information processing mode.

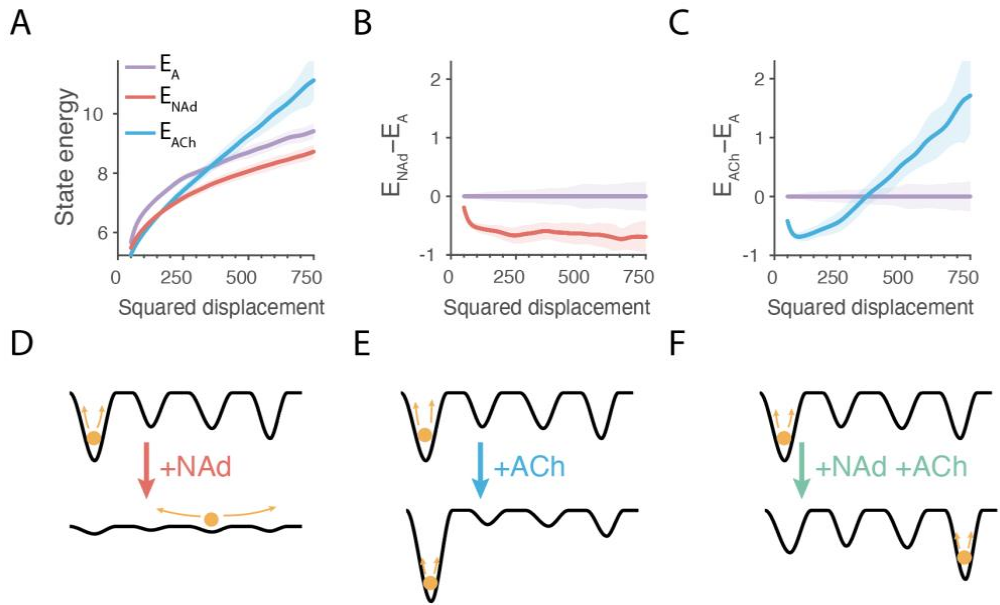

**Figure 5: Energy landscape under low-arousal and adrenergic and cholinergic neuromodulation.** A) Brain-state energy defined as changes in population firing rates, calculated across the low arousal, adrenergic and cholinergic regimes from Fig 4. B) Relative to baseline NAD decreases the energy of all brain state transitions (i.e., makes all state changes more likely). C) Relative to baseline ACH preferences local state (no change) and making large changes high energy relative to baseline. D) Conceptually LC makes it easy to move from one state to another by flattening the landscape. E) ACh deepens energy states, meaning the system is tightly locked within current state or attractor. F) NAd and ACh synergistically combine to facilitate novel brain-state transitions.

## References:

[1] J. M. Shine, E. J. Müller, B. Munn, J. Cabral, R. J. Moran, and M. Breakspear, *Computational Models Link Cellular Mechanisms of Neuromodulation to Large-Scale Neural Dynamics*, Nat Neurosci (2021).
[2] T. W. Schmitz and J. Duncan, *Normalization and the Cholinergic Microcircuit: A Unified Basis for Attention*, Trends in Cognitive Sciences **22**, 422 (2018).
[3] G. Aston-Jones and J. D. Cohen, *An Integrative Theory of Locus Coeruleus-Norepinephrine Function: Adaptive Gain and Optimal Performance*, Annu. Rev. Neurosci. **28**, 403 (2005).
[4] B. R. Munn, E. J. Müller, G. Wainstein, and J. M. Shine, *The Ascending Arousal System Shapes Neural Dynamics to Mediate Awareness of Cognitive States*, Nat Commun **12**, 6016 (2021).
[5] M. Suzuki and M. E. Larkum, *General Anesthesia Decouples Cortical Pyramidal Neurons*, Cell **180**, 666 (2020).
[6] C. Labarrera, Y. Deitcher, A. Dudai, B. Weiner, A. Kaduri Amichai, N. Zylbermann, and M. London, *Adrenergic Modulation Regulates the Dendritic Excitability of Layer 5 Pyramidal Neurons In Vivo*, Cell Reports **23**, 1034 (2018).
[7] S. R. Williams and L. N. Fletcher, *A Dendritic Substrate for the Cholinergic Control of Neocortical Output Neurons*, Neuron **101**, 486 (2019).

[8] J. M. Shine, *Neuromodulatory Influences on Integration and Segregation in the Brain*, Trends in Cognitive Sciences **23**, 572 (2019).
[9] M. E. Larkum, J. J. Zhu, and B. Sakmann, *A New Cellular Mechanism for Coupling Inputs Arriving at Different Cortical Layers*, Nature **398**, 338 (1999).
[10] S. R. Williams, *Dependence of EPSP Efficacy on Synapse Location in Neocortical Pyramidal Neurons*, Science **295**, 1907 (2002).
[11] M. Larkum, *A Cellular Mechanism for Cortical Associations: An Organizing Principle for the Cerebral Cortex*, Trends in Neurosciences **36**, 141 (2013).
[12] M. L. S. Tantirigama, T. Zolnik, B. Judkewitz, M. E. Larkum, and R. N. S. Sachdev, *Perspective on the Multiple Pathways to Changing Brain States*, Front. Syst. Neurosci. **14**, 23 (2020).
[13] N. Takahashi, T. G. Oertner, P. Hegemann, and M. E. Larkum, *Active Cortical Dendrites Modulate Perception*, Science **354**, 1587 (2016).
[14] J. Reimer, M. J. McGinley, Y. Liu, C. Rodenkirch, Q. Wang, D. A. McCormick, and A. S. Tolias, *Pupil Fluctuations Track Rapid Changes in Adrenergic and Cholinergic Activity in Cortex*, Nat Commun **7**, 13289 (2016).
[15] D. Kahneman and J. Beatty, *Pupil Diameter and Load on Memory*, Science **154**, 1583 (1966).
[16] E. H. Hess and J. M. Polt, *Pupil Size in Relation to Mental Activity during Simple Problem-Solving*, Science **143**, 1190 (1964).
[17] S. Joshi and J. I. Gold, *Pupil Size as a Window on Neural Substrates of Cognition*, Trends in Cognitive Sciences **24**, 466 (2020).
[18] S. Joshi, Y. Li, R. M. Kalwani, and J. I. Gold, *Relationships between Pupil Diameter and Neuronal Activity in the Locus Coeruleus, Colliculi, and Cingulate Cortex*, Neuron **89**, 221 (2016).
[19] S. Nieuwenhuis, E. J. De Geus, and G. Aston-Jones, *The Anatomical and Functional Relationship between the P3 and Autonomic Components of the Orienting Response: P3 and Orienting Response*, Psychophysiology **48**, 162 (2011).
[20] F. Cazettes, D. Reato, J. P. Morais, A. Renart, and Z. F. Mainen, *Phasic Activation of Dorsal Raphe Serotonergic Neurons Increases Pupil Size*, Current Biology **31**, 192 (2021).
[21] B. N. Mallick, S. Kaur, and R. N. Saxena, *Interactions between Cholinergic and GABAergic Neurotransmitters in and around the Locus Coeruleus for the Induction and Maintenance of Rapid Eye Movement Sleep in Rats*, Neuroscience **104**, 467 (2001).
[22] J. Cornwall, J. D. Cooper, and O. T. Phillipson, *Afferent and Efferent Connections of the Laterodorsal Tegmental Nucleus in the Rat*, Brain Research Bulletin **25**, 271 (1990).
[23] B. Vitiello, *Cognitive and Behavioral Effects of Cholinergic, Dopaminergic, and Serotonergic Blockade in Humans*, Neuropsychopharmacology **16**, 15 (1997).
[24] B. E. Jones, *Activity, Modulation and Role of Basal Forebrain Cholinergic Neurons Innervating the Cerebral Cortex*, in *Progress in Brain Research*, Vol. 145 (Elsevier, 2004), pp. 157–169.
[25] Y. P. Hou, I. D. Manns, and B. E. Jones, *Immunostaining of Cholinergic Pontomesencephalic Neurons for A1 versus A2 Adrenergic Receptors Suggests Different Sleep–Wake State Activities and Roles*, Neuroscience **114**, 517 (2002).
[26] S. Bouret and S. J. Sara, *Network Reset: A Simplified Overarching Theory of Locus Coeruleus Noradrenaline Function*, Trends in Neurosciences **28**, 574 (2005).
[27] H. Bouma and L. C. J. Baghuis, *Hippus of the Pupil: Periods of Slow Oscillations of Unknown Origin*, Vision Research **11**, 1345 (1971).
[28] C. Stringer, M. Pachitariu, N. Steinmetz, C. B. Reddy, M. Carandini, and K. D. Harris, *Spontaneous Behaviors Drive Multidimensional, Brainwide Activity*, Science **364**, eaav7893 (2019).

[29] J. J. Jun, N. A. Steinmetz, J. H. Siegle, D. J. Denman, M. Bauza, B. Barbarits, A. K. Lee, C. A. Anastassiou, A. Andrei, Ç. Aydın, M. Barbic, T. J. Blanche, V. Bonin, J. Couto, B. Dutta, S. L. Gratiy, D. A. Gutnisky, M. Häusser, B. Karsh, P. Ledochowitsch, C. M. Lopez, C. Mitelut, S. Musa, M. Okun, M. Pachitariu, J. Putzeys, P. D. Rich, C. Rossant, W. Sun, K. Svoboda, M. Carandini, K. D. Harris, C. Koch, J. O'Keefe, and T. D. Harris, *Fully Integrated Silicon Probes for High-Density Recording of Neural Activity*, Nature **551**, 232 (2017).

[30] Q. Wang, S.-L. Ding, Y. Li, J. Royall, D. Feng, P. Lesnar, N. Graddis, M. Naeemi, B. Facer, A. Ho, T. Dolbeare, B. Blanchard, N. Dee, W. Wakeman, K. E. Hirokawa, A. Szafer, S. M. Sunkin, S. W. Oh, A. Bernard, J. W. Phillips, M. Hawrylycz, C. Koch, H. Zeng, J. A. Harris, and L. Ng, *The Allen Mouse Brain Common Coordinate Framework: A 3D Reference Atlas*, Cell **181**, 936 (2020).

[31] W. Gerstner, A. K. Kreiter, H. Markram, and A. V. M. Herz, *Neural Codes: Firing Rates and Beyond*, Proceedings of the National Academy of Sciences **94**, 12740 (1997).

[32] D. Attwell and C. Iadecola, *The Neural Basis of Functional Brain Imaging Signals*, Trends in Neurosciences **25**, 621 (2002).

[33] N. K. Logothetis, *The Underpinnings of the BOLD Functional Magnetic Resonance Imaging Signal*, J. Neurosci. **23**, 3963 (2003).

[34] N. K. Logothetis, *What We Can Do and What We Cannot Do with FMRI*, Nature **453**, 869 (2008).

[35] A. Jafarian, V. Litvak, H. Cagnan, K. J. Friston, and P. Zeidman, *Comparing Dynamic Causal Models of Neurovascular Coupling with FMRI and EEG/MEG*, NeuroImage **216**, 116734 (2020).

[36] K. J. Miller, *Broadband Spectral Change: Evidence for a Macroscale Correlate of Population Firing Rate?*, Journal of Neuroscience **30**, 6477 (2010).

[37] J. R. Manning, J. Jacobs, I. Fried, and M. J. Kahana, *Broadband Shifts in Local Field Potential Power Spectra Are Correlated with Single-Neuron Spiking in Humans*, Journal of Neuroscience **29**, 13613 (2009).

[38] K. Whittingstall and N. K. Logothetis, *Frequency-Band Coupling in Surface EEG Reflects Spiking Activity in Monkey Visual Cortex*, Neuron **64**, 281 (2009).

[39] Ö. Yüzgeç, M. Prsa, R. Zimmermann, and D. Huber, *Pupil Size Coupling to Cortical States Protects the Stability of Deep Sleep via Parasympathetic Modulation*, Current Biology **28**, 392 (2018).

[40] A. Renart, J. de la Rocha, P. Bartho, L. Hollender, N. Parga, A. Reyes, and K. D. Harris, *The Asynchronous State in Cortical Circuits*, Science **327**, 587 (2010).

[41] E. Samuels and E. Szabadi, *Functional Neuroanatomy of the Noradrenergic Locus Coeruleus: Its Roles in the Regulation of Arousal and Autonomic Function Part II: Physiological and Pharmacological Manipulations and Pathological Alterations of Locus Coeruleus Activity in Humans*, CN **6**, 254 (2008).

[42] L. Zaborszky, L. Hoemke, H. Mohlberg, A. Schleicher, K. Amunts, and K. Zilles, *Stereotaxic Probabilistic Maps of the Magnocellular Cell Groups in Human Basal Forebrain*, NeuroImage **42**, 1127 (2008).

[43] S. X. Chen, A. N. Kim, A. J. Peters, and T. Komiyama, *Subtype-Specific Plasticity of Inhibitory Circuits in Motor Cortex during Motor Learning*, Nat Neurosci **18**, 1109 (2015).

[44] M. E. Larkum, T. Nevian, M. Sandler, A. Polsky, and J. Schiller, *Synaptic Integration in Tuft Dendrites of Layer 5 Pyramidal Neurons: A New Unifying Principle*, Science **325**, 756 (2009).

[45] M. J. Redinbaugh, J. M. Phillips, N. A. Kambi, S. Mohanta, S. Andryk, G. L. Dooley, M. Afrasiabi, A. Raz, and Y. B. Saalmann, *Thalamus Modulates Consciousness via Layer-Specific Control of Cortex*, Neuron S0896627320300052 (2020).

[46] J. M. Beggs and D. Plenz, *Neuronal Avalanches in Neocortical Circuits*, J. Neurosci. **23**, 11167 (2003).

[47] M. E. Hasselmo, *The Role of Acetylcholine in Learning and Memory*, Current Opinion in Neurobiology **16**, 710 (2006).

[48] A. Reina, T. Bose, V. Trianni, and J. A. R. Marshall, *Psychophysical Laws and the Superorganism*, Sci Rep **8**, 4387 (2018).

[49] O. Kinouchi and M. Copelli, *Optimal Dynamical Range of Excitable Networks at Criticality*, Nature Phys **2**, 348 (2006).

[50] L. L. Gollo, *Coexistence of Critical Sensitivity and Subcritical Specificity Can Yield Optimal Population Coding*, J. R. Soc. Interface **14**, 20170207 (2017).

[51] J. P. Cunningham and B. M. Yu, *Dimensionality Reduction for Large-Scale Neural Recordings*, Nature Neuroscience **17**, 11 (2014).

[52] M. Jazayeri and A. Afraz, *Navigating the Neural Space in Search of the Neural Code*, Neuron **93**, 1003 (2017).

[53] Y. J. John, K. S. Sawyer, K. Srinivasan, E. J. Müller, B. R. Munn, and J. M. Shine, *It's about Time: Linking Dynamical Systems with Human Neuroimaging to Understand the Brain*, Network Neuroscience 1 (2022).

[54] J. M. Shine, M. J. Aburn, M. Breakspear, and R. A. Poldrack, *The Modulation of Neural Gain Facilitates a Transition between Functional Segregation and Integration in the Brain*, ELife **7**, e31130 (2018).

[55] D. Krzemiński, N. Masuda, K. Hamandi, K. D. Singh, B. Routley, and J. Zhang, *Energy Landscape of Resting Magnetoencephalography Reveals Fronto-Parietal Network Impairments in Epilepsy*, Network Neuroscience **4**, 374 (2020).

[56] G. Tkačik, T. Mora, O. Marre, D. Amodei, S. E. Palmer, M. J. Berry, and W. Bialek, *Thermodynamics and Signatures of Criticality in a Network of Neurons*, Proc Natl Acad Sci USA **112**, 11508 (2015).

[57] W. Bialek, *Perspectives on Theory at the Interface of Physics and Biology*, Rep. Prog. Phys. **81**, 012601 (2018).

[58] B. Munn and P. Gong, *Critical Dynamics of Natural Time-Varying Images*, Phys. Rev. Lett. **121**, 058101 (2018).

[59] M. E. Hasselmo and J. McGaughy, *High Acetylcholine Levels Set Circuit Dynamics for Attention and Encoding and Low Acetylcholine Levels Set Dynamics for Consolidation*, in *Progress in Brain Research*, Vol. 145 (Elsevier, 2004), pp. 207–231.

[60] A. S. Ecker, P. Berens, R. J. Cotton, M. Subramaniyan, G. H. Denfield, C. R. Cadwell, S. M. Smirnakis, M. Bethge, and A. S. Tolias, *State Dependence of Noise Correlations in Macaque Primary Visual Cortex*, Neuron **82**, 235 (2014).

[61] M. M. Churchland, B. M. Yu, J. P. Cunningham, L. P. Sugrue, M. R. Cohen, G. S. Corrado, W. T. Newsome, A. M. Clark, P. Hosseini, B. B. Scott, D. C. Bradley, M. A. Smith, A. Kohn, J. A. Movshon, K. M. Armstrong, T. Moore, S. W. Chang, L. H. Snyder, S. G. Lisberger, N. J. Priebe, I. M. Finn, D. Ferster, S. I. Ryu, G. Santhanam, M. Sahani, and K. V. Shenoy, *Stimulus Onset Quenches Neural Variability: A Widespread Cortical Phenomenon*, Nat Neurosci **13**, 369 (2010).

[62] M. Jing, Y. Li, J. Zeng, P. Huang, M. Skirzewski, O. Kljakic, W. Peng, T. Qian, K. Tan, J. Zou, S. Trinh, R. Wu, S. Zhang, S. Pan, S. A. Hires, M. Xu, H. Li, L. M. Saksida, V. F. Prado, T. J. Bussey, M. A. M. Prado, L. Chen, H. Cheng, and Y. Li, *An Optimized Acetylcholine Sensor for Monitoring in Vivo Cholinergic Activity*, Nat Methods **17**, 1139 (2020).

[63] J. Feng, C. Zhang, J. E. Lischinsky, M. Jing, J. Zhou, H. Wang, Y. Zhang, A. Dong, Z. Wu, H. Wu, W. Chen, P. Zhang, J. Zou, S. A. Hires, J. J. Zhu, G. Cui, D. Lin, J. Du, and Y. Li, *A Genetically Encoded Fluorescent Sensor for Rapid and Specific In Vivo Detection of Norepinephrine*, Neuron **102**, 745 (2019).

[64] M. E. Larkum, *Top-down Dendritic Input Increases the Gain of Layer 5 Pyramidal Neurons*, Cerebral Cortex **14**, 1059 (2004).

[65] Z. Wang and D. McCormick, *Control of Firing Mode of Corticotectal and Corticopontine Layer V Burst-Generating Neurons by Norepinephrine, Acetylcholine, and 1S,3R- ACPD*, J. Neurosci. **13**, 2199 (1993).

[66] K. Grzelka, P. Kurowski, M. Gawlak, and P. Szulczyk, *Noradrenaline Modulates the Membrane Potential and Holding Current of Medial Prefrontal Cortex Pyramidal Neurons via B1-Adrenergic Receptors and HCN Channels*, Front. Cell. Neurosci. **11**, 341 (2017).

[67] E. M. Izhikevich, *Simple Model of Spiking Neurons*, IEEE Transactions on Neural Networks **14**, 1569 (2003).

[68] J. Szentágothai, *The 'Module-Concept' in Cerebral Cortex Architecture*, Brain Research **95**, 475 (1975).

[69] S. Heitmann, T. Boonstra, and M. Breakspear, *A Dendritic Mechanism for Decoding Traveling Waves: Principles and Applications to Motor Cortex*, PLoS Comput Biol **9**, e1003260 (2013).

[70] B. W. Connors and M. J. Gutnick, *Intrinsic Firing Patterns of Diverse Neocortical Neurons*, Trends in Neurosciences **13**, 99 (1990).

[71] M. Wang, B. P. Ramos, C. D. Paspalas, Y. Shu, A. Simen, A. Duque, S. Vijayraghavan, A. Brennan, A. Dudley, E. Nou, J. A. Mazer, D. A. McCormick, and A. F. T. Arnsten, *A2A-Adrenoceptors Strengthen Working Memory Networks by Inhibiting CAMP-HCN Channel Signaling in Prefrontal Cortex*, Cell **129**, 397 (2007).

[72] S. H. Gautam, T. T. Hoang, K. McClanahan, S. K. Grady, and W. L. Shew, *Maximizing Sensory Dynamic Range by Tuning the Cortical State to Criticality*, PLoS Comput Biol **11**, e1004576 (2015).

[73] W. L. Shew, H. Yang, T. Petermann, R. Roy, and D. Plenz, *Neuronal Avalanches Imply Maximum Dynamic Range in Cortical Networks at Criticality*, Journal of Neuroscience **29**, 15595 (2009).

[74] S. S. Stevens, *On the Psychophysical Law.*, Psychological Review **64**, 153 (1957).